\documentclass[12pt,preprint]{aastex}

\slugcomment{}

\begin{document}

\title{
From primordial $^4$He abundance to the Higgs field.}


\author{Josef M. Ga{\ss}ner, Harald Lesch}
\affil{University Observatory Munich, Scheinerstr. 1, 81679 Munich,
Germany}

\and

\author{Hartmuth Arenh\"{o}vel}
\affil{Institut f\"{u}r Kernphysik, Johannes Gutenberg-Universit\"{a}t, 55099 Mainz, Germany}

\begin{abstract}
We constrain the possible time variation of the Higgs vacuum expectation value ($v$) by
recent results on the primordial $^4$He abundance ($Y_P$). For that, we improve the analytic models of the key-processes in our previous analytic calculation of the primordial $^4$He abundance. Furthermore, the latest results on the neutron decay, the baryon to photon ratio based on 5-year WMAP observations and a new dependence of the deuteron binding energy on $v$ are incorporated.

Finally, we approximate the weak freeze-out, the cross section of photo-disintegration of the deuteron, the mean lifetime of the free neutron, the mass difference of neutron and proton, the Fermi coupling constant, the mass of the electron and the binding energy of the deuteron by terms of $v$, to constrain its possible time variation by recent results on the primordial $^4$He abundance:
$\left|\frac{\Delta v}{v}\right| ~ \leq 1.5 \cdot 10^{-4}$.
\end{abstract}


\section{Introduction}
The standard model \citep{Griffiths} is a remarkably successful description of
fundamental particle interactions.  The theory contains parameters - such as particle
masses - whose origins are still unknown and which cannot be predicted, but whose values
are constrained through their interactions with the conjectured Higgs field. The Higgs
field is assumed to have a non-zero value in the ground state of the universe - called
its vacuum expectation value $v$ - and elementary particles that interact with the Higgs
field obtain a mass
proportional to this fundamental constant of nature. \\
Although the question whether the fundamental constants are in fact constant, has a long
history of study (see \cite{Uzan} for a review), comparatively less interest has been directed towards the consequences of a possible variation of $v$ \citep{Gas08,Den08,Cha07,Lan06,Li06,Yoo,
Ichikawa, Kujat,Scherrer,Dixit}.\\
A macroscopic probe to determine the allowed variation range is given by the network of
nuclear interactions during the Big-Bang-Nucleosynthesis (see \cite{Yao06} for a
review of the Standard Big-Bang-Nucleosynthesis Model SBBN), with its final primordial abundance of
$^4$He.
The relevant key-parameters are the freeze-out concentration of neutrons and
protons, the so called deuterium bottleneck (the effective start of the primordial
nucleosynthesis) and the neutron decay. Their dependency on $v$ and the final impact on the resulting primordial abundance of $^4$He can be understood more clearly by an indepth approach.

Here we present a revised calculation of the primordial $^4$He abundance \citep{Gas08}, where the analytic models of all key-processes have been improved.
The opening of the deuterium bottleneck and the weak freeze-out are determined more accurately and a new dependence of the deuteron binding energy on $v$ is incorporated, based on different nucleon-nucleon-potential-models.
\\
The analytic approach enables us to take important issues into consideration, that have been ignored
by previous authors, as there are the $v$-dependence of the relevant cross sections of deuteron
production and its photo-disintegration. Furthermore, we take a non-equilibrium Ansatz for the freeze-out concentration of neutrons and protons and incorporate the latest results on the neutron decay and the baryon to photon ratio.

Finally, we approximate the weak freeze-out, the cross section of photo-disintegration of the deuteron, the mean lifetime of the free neutron, the mass difference of neutron and proton, the Fermi coupling constant, the mass of the electron and the binding energy of the deuteron by terms of $v$, to constrain its possible time variation by recent results on the primordial $^4$He abundance \citep{Pei07,Izo07}.

We briefly note, that constraints on the spacial variation of $v$ require a measurement
of helium abundance anisotropy or inhomogeneity versus the position in the sky and an
inhomogeneous theoretical BBN model. The homogeneous formalism used throughout the paper
thus assumes a spacial invariance of the Higgs vacuum expectation value.

\section{Calculations}

All relevant processes of SBBN took place at a very early epoch, when the energy density
was dominated by radiation, leading to a time-temperature relation for a flat universe:
\begin{eqnarray}\label{t_aus_T} t =
\sqrt{\frac{90 \hbar ^3 c^5}{32 \pi ^3 k^4 G g_*}} \frac{1}{T^2} ~[s],
\end{eqnarray}
where $c$ is the velocity of light, $k$ the Boltzmann constant, $G$ denotes the
gravitational constant and $\hbar$ is the Planck constant divided by 2$\pi$. $g_*$ counts
the total number of effectively massless ($mc^2 \ll kT$) degrees of freedom, given by
$g_* = \left( g_b + \frac{7}{8}g_f \right)$, in which $g_b$ represents the bosonic and
$g_f$ the fermionic contributions at the relevant temperature.
\\
At very high temperatures ($T\gg 10^{10}$K), the neutrons and protons are kept in thermal
and chemical equilibrium by the weak interactions
\begin{eqnarray*}
n + e^+ &\rightleftarrows&  p + \bar{\nu}_e, \\
n + \nu_e  &\rightleftarrows&  p + e^- ~ { and }\\
n  &\rightleftarrows&  p + e^- + \bar{\nu}_e,
\end{eqnarray*}
until the temperature drops to a certain level, at which the inverse reactions become
inefficient. This so called "freeze-out"-temperature $T_f$ and time $t_f$ denote the
start of the effective neutron beta decay.

Assuming chemical and thermal equilibrium, the rate of neutron to proton concentration at freeze-out is commonly calculated, assuming chemical and thermal equilibrium:
\begin{eqnarray}\label{n_n_durch_np_GG}
\frac{n_n}{n_p}(T_f) = e^{-\frac{Q}{kT_f}},
\end{eqnarray}
where Q denotes the energy difference of neutron and proton rest masses.
However, the deviation from equilibrium at freeze-out is already significant. Hence, we have to derive non-equilibrium concentrations, where we follow the example calculations of \cite{Muk04}:\\
The 4-fermion-interaction $a + b \rightarrow  c + d$ can be calculated using the Fermi theory, where the differential cross section is given by
\begin{eqnarray}\label{fermi_WW}
\frac{d \sigma_{ab}}{d\Omega} = \frac{1}{(8 \pi)^2} \frac{|\emph{M} |^2}{(p_a+p_b)^2} \sqrt{ \frac{(p_c \cdot p_d)^2 - m_c^2~ m_d^2~ c^8}{(p_a \cdot p_b)^2 - m_a^2~ m_b^2~ c^8}}.
\end{eqnarray}
$(p_a \cdot p_b)$ and $(p_c \cdot p_d)$ denote the scalar products of the 4-momenta and the matrix element is given by:
\begin{eqnarray*}
|\emph{M} |^2 = 16 (1+3g_A^2)~G_F^2~(p_a \cdot p_b)(p_c \cdot p_d),
\end{eqnarray*}
where $G_F \simeq 1.166371\cdot 10^{-11}~[\frac{1}{MeV^2}]$ \citep{Yao06} denotes the Fermi coupling constant and $g_A=1.2739$ \citep{Abe02} the axial vector coupling constant, respectively.

First, we consider the reaction $n + \nu_e \rightarrow  p + e^-$ at the relevant temperatures around few MeV and below, where the nucleons are nonrelativistic:
\begin{eqnarray*}
(p_n + p_\nu)^2 &\simeq& m_n^2 c^4\\
(p_n \cdot p_\nu) &=& m_n c^2~\epsilon_\nu\\
(p_p \cdot p_e)&=& m_p c^2~ \epsilon_e\\
\sqrt{(p_p \cdot p_e)^2 - m_p^2 m_e^2 c^8} &\simeq& m_p~ c^2~ \epsilon_e ~\sqrt{1 - \left(\frac{m_e c^2}{\epsilon_e}\right)^2} = m_p ~c~\epsilon_e ~v_e
\end{eqnarray*}
where $m_p$, $m_n$ and $m_e$ denote the mass of the proton, neutron and electron, respectively, $v_e$ is the velocity of the electron, $\epsilon_\nu$ is the energy of the incoming neutrino and $\epsilon_e \simeq \epsilon_\nu + Q$ is the energy of the outgoing electron. Substituting all terms into eq. (\ref{fermi_WW}) we obtain:
\begin{eqnarray}
\frac{d \sigma_{n\nu}}{d\Omega} &=& \frac{1}{(8 \pi)^2}~ 16 ~(1+3g_A^2)~G_F^2 ~\frac{m_n \epsilon_\nu m_p \epsilon_e}{m_n^2} \frac{m_p \epsilon_e \frac{v_e}{c}}{\sqrt{(m_n \epsilon_\nu)^2 - (m_n m_\nu c^2)^2}}\\
&=& \frac{1}{(2 \pi)^2}~ (1+3g_A^2)~G_F^2 ~\frac{m_p^2}{m_n^2} ~\epsilon_e^2 ~\frac{v_e}{c},
\end{eqnarray}
where we neglect the neutrino mass $m_\nu$. Integration leads to
\begin{eqnarray}
\sigma_{n\nu}= \frac{1+3g_A^2}{\pi}~ G_F^2~ \frac{m_p^2}{m_n^2}~ \epsilon_e^2~ \frac{v_e}{c}.
\end{eqnarray}
Next, we have to consider that at temperatures $kT > 2~m_e c^2$ the possible states for the electron are partially occupied by electron-positron-pairs. According to the Pauli exclusion principle, this reduces the appropriate cross section to
\begin{eqnarray}\label{sigma_nnu}
\sigma_{n\nu}^*= \sigma_{n\nu}~ \frac{1}{1+ e^{-\frac{\epsilon_e}{kT}}}.
\end{eqnarray}
This enables us to calculate $\Delta N_n$, the reduction of neutrons within a time interval $\Delta t$ in a given volume, containing $N_n$ neutrons:
\begin{eqnarray}\label{DeltaN_n}
\Delta N_n = - \left( \sum_{\epsilon_\nu}^{\epsilon_\nu+\Delta\epsilon_\nu} \sigma_{n\nu}^* n_{\epsilon_\nu} v_\nu \Delta g_{\epsilon_\nu} \right) N_n \Delta t. \end{eqnarray}
where
\begin{eqnarray}
     n_{\epsilon_\nu} = \frac{1}{1+e^{\frac{\epsilon_\nu}{kT_\nu}}}
\end{eqnarray}
denotes the neutrino occupation number ($v_\nu$ and $T_\nu$ are the velocity and the temperature of the neutrinos) and
\begin{eqnarray}
     \Delta g_{\epsilon_\nu} = \frac{1}{2\pi^2} \int_{\epsilon_\nu}^{\epsilon_\nu + \Delta \epsilon_\nu} |p|^2 ~d|p| \simeq \frac{1}{2\pi}~ \epsilon_\nu^2 ~\Delta \epsilon_\nu
\end{eqnarray}
the phase volume element. Introducing the relative concentration of the neutrons
\begin{eqnarray}
X_n =
\frac{n_n}{n_n + n_p}
\end{eqnarray}
and assuming baryon conservation, we obtain the rate of change of the neutron concentration due to the $n\nu$-process:
\begin{eqnarray}
\left(\frac{d X_n}{dt} \right)_{n\nu} = - \lambda_{n\nu} X_n,
\end{eqnarray}
where $\lambda_{n\nu}$ denotes the decay rate.
Substituting the cross section (\ref{sigma_nnu}) into eq. (\ref{DeltaN_n}) we obtain:
\begin{eqnarray}
\lambda_{n\nu}=\frac{1+3g_A^2}{2\pi^3}~ G_F^2 ~\frac{m_p^2}{m_n^2}~\mathfrak{I}(T_\nu),
\end{eqnarray}
where
\begin{eqnarray}
\mathfrak{I}(T_\nu) = \int_0^\infty
\epsilon_e^2~\sqrt{1-\left(\frac{m_e c^2}{\epsilon_e}\right)^2}~\frac{1}{1+e^{-\frac{\epsilon_e}{kT}}} ~\frac{\epsilon_\nu^2}{1+e^{\frac{\epsilon_\nu}{kT_\nu}}}~ d\epsilon_\nu.
\end{eqnarray}

Below the temperature $kT \simeq 2~m_e ~c^2$, the Pauli exclusion principle, represented by the term $(1+e^{-\frac{\epsilon_e}{kT}})$, looses importance and numerically we notice a deviation of 1 \% only, when we set this term to 1. Expanding the square root ($m_e c^2 / \epsilon_e  \ll 1$) keeping only first two terms and introducing the integration variable $x = \frac{\epsilon_\nu}{kT_\nu}$ we derive:
\begin{eqnarray}
\mathfrak{I}(T_\nu) &=& (kT_\nu)^5 \int_0^\infty x^2 \frac{\left(x+\frac{Q}{kT_\nu}\right)^2-\frac{1}{2}\left(\frac{m_e c^2}{kT_\nu}\right)^2}{1+e^x}~ dx \\ \nonumber \\
&=& Q^5~\left( \frac{kT_\nu}{Q}\right)^3 \left[ \frac{45 ~\zeta (5)}{2} \left( \frac{kT_\nu}{Q}\right)^2 + \frac{7 \pi^4}{60} \left( \frac{kT_\nu}{Q}\right) + \frac{3 ~\zeta (3)}{2} \left( 1-\frac{m_e^2c^4}{2~Q^2} \right) \right]\\ \nonumber \\
&\simeq& Q^5~\frac{45 ~\zeta (5)}{2}~\left( \frac{kT_\nu}{Q}\right)^3 \left( \frac{kT_\nu}{Q}+ 0.25 \right)^2,
\end{eqnarray}
where $\zeta$ is the Riemann zeta function. In the last step, we completed the square approximately. Finally we convert $\lambda_{n\nu}$ from MeV to $\frac{1}{s}$ and derive
\begin{eqnarray}\label{lambda}
\lambda_{n\nu} \simeq \frac{1+3g_A^2}{1.75 \cdot 10^{-21}}~ G_F^2 ~\frac{m_p^2}{m_n^2}~Q^5 \left( \frac{kT_\nu}{Q}\right)^3 \left( \frac{kT_\nu}{Q}+ 0.25 \right)^2 ~\left[\frac{1}{s} \right].
\end{eqnarray}
Similarly, we find the decay rate of the reaction $n + e^+ \rightarrow p + \bar{\nu}$ (we interchange $\epsilon_\nu$ with $\epsilon_e$ and $m_e$ with $m_\nu = 0$):
\begin{eqnarray}
\lambda_{ne}=\frac{1+3g_A^2}{2\pi^3}~ G_F^2 ~\frac{m_p^2}{m_n^2}~\int_{m_ec^2}^{\infty}
\epsilon_\nu^2~\frac{\epsilon_e^2}{1+e^{\frac{\epsilon_e}{kT}}}~ d\epsilon_e
\end{eqnarray}
Assuming $T_\nu = T$, the rates of the inverse reactions are related to the rate of the direct reactions as
\begin{eqnarray}
\lambda_{pe}=e^{-\frac{Q}{kT}} \lambda_{n\nu}\\
\lambda_{p\nu}=e^{-\frac{Q}{kT}} \lambda_{ne}.
\end{eqnarray}
Hence, we can write the following balance equation for $X_n$:
\begin{eqnarray}\label{freeze-out}
\frac{d X_n}{dt} &=& - (\lambda_{n\nu} + \lambda_{ne}) X_n + (\lambda_{pe} + \lambda_{p\nu}) (1-X_n) \nonumber \\
& = &- (\lambda_{n\nu} + \lambda_{ne}) (1+e^{-\frac{Q}{kT}})(X_n - X_n^{eq})
\end{eqnarray}
with the equilibrium neutron concentration
\begin{eqnarray}
 X_n^{eq} = \frac{1}{1+e^{\frac{Q}{kT}}}.
\end{eqnarray}
To solve this linear differential equation (\ref{freeze-out}), we take the initial condition $X_n(t=0) = X_n^{eq}$ and obtain:
\begin{eqnarray}\label{X_n_DGL}
 X_n(t) =  X_n^{eq}(t) - \int_0^t \exp{ \left( - \int_{\tilde{t}}^t (\lambda_{n\nu}(y) + \lambda_{ne}(y)) (1+e^{-\frac{Q}{kT}})dy \right)} \dot{X}_n^{eq}(\tilde{t}) d\tilde{t},
\end{eqnarray}
where dot denotes the derivative with respect to time. Using the auxiliary function $F(t)$
\begin{eqnarray}
F(t) =  \int_0^t (\lambda_{n\nu}(t) + \lambda_{ne}(t)) (1+e^{-\frac{Q}{kT}})dt,
\end{eqnarray}
we express the integral in (\ref{X_n_DGL}) in the form
\begin{eqnarray}
     \int_0^t e^{-F(t) + F(\tilde{t})}~\dot{X}_n^{eq}(\tilde{t})~     d\tilde{t}
\end{eqnarray}
and expand in the small parameter $(t-\tilde{t})$, since the integral is dominated by the contribution of $\tilde{t}\simeq t$ if $F(t)$ is a quickly growing function of $t$:
\begin{eqnarray}
X_n(t) =  X_n^{eq}(t) - \int_0^t \left[ \dot{X}_n^{eq}(t)+\ddot{X}_n^{eq}(t)(\tilde{t}-t) +...\right] ~e^{-\dot{F}(t-\tilde{t})} \left( 1 + \frac{1}{2} \ddot{F}(t) (\tilde{t}-t)^2 +... \right) dt
\end{eqnarray}
We integrate term by term using
\begin{eqnarray}
     \int_0^t e^{-A(t-\tilde{t})} (t-\tilde{t})^n~ d\tilde{t}\simeq A^{-n-1}~n!,
\end{eqnarray}
where we neglect exponentially small terms of order $e^{-At}$, deriving:
\begin{eqnarray}\label{X_n_2}
 X_n(t) =  X_n^{eq}(t) \left( 1 - \frac{1}{(\lambda_{n\nu} + \lambda_{ne}) (1+e^{-\frac{Q}{kT}})} \frac{{\dot{X}}_n^{eq}(t)}{X_n^{eq}(t)} +... \right).
\end{eqnarray}
Later, when the temperature has dropped significantly, $X_n^{eq}$ goes to zero and the integral in eq. (\ref{X_n_DGL}) approches the finite limit. As a result, the neutron concentration freezes-out at $X_n(t\rightarrow \infty)$. Effectively, this freeze-out occurs, when the deviation from equilibrium becomes significant, hence when
\begin{eqnarray}\label{X_n_3}
\frac{\dot{X}_n^{eq}}{X_n^{eq}} \simeq (\lambda_{n\nu} + \lambda_{ne}) (1+e^{-\frac{Q}{kT}}).
\end{eqnarray}
Assuming this happens before $e^{\pm}$-annihilation and after $kT$ has dropped below $Q$, we set $\lambda_{n\nu} + \lambda_{ne} \simeq 2 \lambda_{n\nu}$ and neglect the term $\exp(-Q/kT)$. Substituting all terms and taking the time-temperatur-relation (\ref{t_aus_T}) into account, we finally obtain an equation for the freeze-out-temperature $T_f$:
\begin{eqnarray}\label{T_f}
\sqrt{\frac{G g_*(T_f)}{\hbar^3c^5}}~3.7 \cdot 10^{-35} = (1+3g_A^2)~G_F^2~Q^3~\left( \frac{kT_f}{Q} \right)^2 \left( \frac{kT_f}{Q}+ 0.25 \right)^2.
\end{eqnarray}
The quadratic term $\frac{kT_f}{Q}$ leads to
\begin{eqnarray}
T_f \simeq 1.16 \cdot 10^6~\frac{Q}{k}~ \left[-\frac{1}{8}+\sqrt{\frac{1}{64}+\sqrt{\frac{3.7 \cdot 10^{-35}}{(1+3g_A^2)~G_F^2~Q^3}     }\left(\frac{G g_*(T_f)}{\hbar^3c^5}\right)^{1/4}} \right]~[K].
\end{eqnarray}
At $T_f$ the effectively massless species in the cosmic plasma are neutrinos (left-handed only), antineutrinos (right-handed only), electrons, positrons and photons. For the case of three neutrino families ($N_\nu=3)$, we obtain $g_*(T_f) = (2 + \frac{7}{8} (4+2 N_\nu) = 10.75$.\\
To calculate the relevant neutron concentration at $T_f$, we go back to eq. (\ref{X_n_DGL}). Since $X_n^{eq}\rightarrow 0$ as $T \rightarrow 0$ we have to calculate the integral term in eq. (\ref{X_n_DGL}) in the limit $t \rightarrow \infty$. The main contribution to the integral comes at temperature above the restmass of the electron, where again $\lambda_{n\nu}+\lambda_{ne} \simeq 2~\lambda_{n\nu}$ ($\lambda_{n\nu}$ given by eq. (\ref{lambda})).
Furthermore, we use eq. (\ref{t_aus_T}) to change the integration variable from $dt$ to $dT$:
\begin{eqnarray}\label{X_n}
X_n(T_f) = \int_0^\infty \frac{Q~\exp\left[-2.68 \cdot 10^{34} \sqrt{\frac{\hbar^3c^5}{G g_*(T_f)}}~ (1+3g_A^2)~G_F^2  \int_0^{T} (x + \frac{Q}{4})^2(1+e^{-\frac{Q}{x}})dx\right] }{2~
T^2 \left(1+\cosh\frac{Q}{T}\right)} dT,
\end{eqnarray}
with $Q$ and $T$ in units of MeV. For later purposes, we finally state the neutron to proton ratio at freeze-out:
\begin{eqnarray}\label{nicht_gg}
\frac{n_n}{n_p}(T_f) = \frac{1}{ \frac{1}{X_n(T_f)} - 1}
\end{eqnarray}
In comparison, the equilibrium rate eq. (\ref{n_n_durch_np_GG}) is 14 \% higher, thus as mentioned before, the deviation is significant, which justifies the effort.

From now on the loss of free neutrons via $n \rightarrow p + e^-
+ \bar{\nu}_e$, with a mean lifetime \citep{Serebrov} $\tau_n=878.5$ s, can no longer
be compensated. Thus, whereas the neutron density decreases as $n_n(t) = n_n(t_f) \cdot
e^{-\frac{t-t_f}{\tau_n}}$, the proton density increases as $n_p(t) = n_p(t_f) +
(n_n(t_f) - n_n(t))$ and we obtain
\begin{eqnarray}\label{n_n/n_p}
\frac{n_p}{n_n}(t) &=& \frac{e^{\frac{t-t_f}{\tau_n}}}{X_n(T_f)} - 1.
\end{eqnarray}

At the relevant densities in the early universe, fusion reactions can only proceed efficiently through sequences of two-body-collisions and the starting product of these collisions is the weakly bound deuteron ($B_d\simeq 2.225 $ MeV), which is highly affected by photo-disintegration. The start of nucleosynthesis, $t_N$, is therefore usually referred to as the "deuterium bottleneck".

Once the deuteron production dominates the photo-disintegration and the expansion of the universe, our calculation in some sense "produces only deuteron", disregarding that deuteron is also destroyed by the fusion of light elements. In fact, we do not consider the detailed fusion reactions with their intermediate products that finally lead to $^4$He. We are interested in the point $t_N$, from that onwards we can assume neutron conservation, because enough neutrons have reached stable states inside light nuclei (no matter whether inside deuterons or further fusion products of deuteron).
Hence, we obtain $t_N$ assuming two constraints: First, the deuteron production must dominate the photo-disintegration and the expansion of the universe. Second, to justify neutron conservation, the deuteron density must exceed the density of free neutrons. In fact, it turns out, that we can reproduce the numerical as well as the observational results for $Y_P$, assuming neutron conservation when $52 - 66$ \% of the neutrons have reached stable states inside light nuclei.\\
The interval between $t_f$ and $t_N$ is a substantial fraction of the neutron lifetime and therefore plays an essential role for the outcome of the primordial helium production.
Thus, we have to calculate the rates of deuteron production $\Gamma_{(np\rightarrow
d\gamma)}$, deuteron photo-disintegration $\Gamma_{(\gamma d\rightarrow np)}$ and the
rate of deuteron density reduction by the expansion of the universe $\Gamma_{expansion}$, to determine $t_N$ respectively $T_N$, when
\begin{eqnarray}\label{gammas}
\Gamma_{(np\rightarrow d\gamma)} > \Gamma_{(\gamma d\rightarrow np)} +\Gamma_{expansion}.
\end{eqnarray}
The rates for production and photo-disintegration of deuteron are given by the product of
the relevant number densities, velocities and cross sections, whereas the expansion rate of a radiation-dominated, flat universe is given by $1 \over{2 t}$ with
t from eq. (\ref{t_aus_T}), leading to:
\begin{eqnarray}
n_n~\frac{\eta~n_\gamma}{1+\frac{n_n}{n_p}}~ \sqrt{\frac{8 k T}{\pi m_N}}~ \sigma_{(np \rightarrow d\gamma)} > n_d~n_{\gamma}^* ~c
~\sigma_{(\gamma d \rightarrow np)} + \frac{n_d}{2 ~t},
\end{eqnarray}
where $\sigma_{(\gamma d \rightarrow np)}$ denotes the cross section of deuteron photo-disintegration, $\sigma_{(np \rightarrow d\gamma)}$ the cross section of deuteron production, $m_N$ is the nucleon mass, $\eta\simeq 6.226 \cdot 10^{-10}$ is the baryon to photon ratio based on WMAP \citep{Kom08} and \cite{Ste06}, $n_d$ and $n_\gamma$ denote the number densities of deuterons and photons, respectively, and $n_{\gamma}^*$ the number density of photons which supply enough energy to disintegrate the deuteron and do not loose this energy in much more likely Compton scattering on electrons.\\
The number density of photons at a certain temperature T is given by
\begin{eqnarray}
n_{\gamma}= \frac{8 \pi}{(h c)^3} \int_0^\infty \frac{E_\gamma^2}{e^{\frac{E_\gamma}{k T}}-1} dE_\gamma = 16
\pi~ \zeta(3) \left(\frac{kT}{hc}\right)^3,
\end{eqnarray}
where $\zeta$ is the Riemann zeta function. The number density of these photons supplying a minimum energy $E_\gamma > B_d \gg kT$ is
\begin{eqnarray}
n_{(\gamma>B_d)} = \frac{8 \pi}{(h c)^3} \int_{B_d}^\infty E_\gamma^2 e^{-\frac{E_\gamma}{k
T}}dE_\gamma = 8 \pi \left(\frac{kT}{hc}\right)^3
\left[\left(\frac{B_d}{kT}+1\right)^2+1\right] e^{-\frac{B_d}{kT}},
\end{eqnarray}
but most of them will loose energy in Compton scattering on electrons, leading to
\begin{eqnarray}
n_{\gamma}^* = n_{(\gamma>B_d)}~ \frac{n_d~\sigma_{(\gamma d \rightarrow np)}}{n_p~\sigma_{(\gamma e \rightarrow  e \gamma)}},
\end{eqnarray}
where $\sigma_{(\gamma e \rightarrow e \gamma)}$ denotes the Klein-Nishina cross section \citep{Rybicki} for Compton scattering on electrons:
\begin{eqnarray}\label{Klein_Nishina}
\sigma_{(\gamma e \rightarrow e \gamma)}= \frac{1}{8 \pi} \left(\frac{e^2}{\epsilon_0 m_e c^2}\right)^2\left[ \frac{1+\beta}{\beta^2} \left( \frac{2 (1+\beta)}{1+2\beta}-\frac{\ln{(1+2\beta)}}{\beta} \right) + \frac{\ln{(1+2\beta)}}{2 \beta} - \frac{1 + 3\beta}{(1+2\beta)^2} \right]
\end{eqnarray}
where
\begin{eqnarray}
\beta = \frac{\langle E_\gamma \rangle}{m_e c^2},
\end{eqnarray}
and the mean incident photon energy $\langle E_\gamma \rangle$ is given by
\begin{eqnarray}
\langle E_\gamma \rangle &=& \frac{1}{n_{(\gamma>B_d)}}\frac{8 \pi}{(h c)^3} \int_{B_d}^\infty
E_\gamma^3 e^{-\frac{E_\gamma}{k T}} dE_\gamma
= k T \left[ \frac{\left(\frac{B_d}{k T}\right)^3}{\left(\frac{B_d}{k T}+1\right)^2+1}+3\right]
~\simeq~ B_d + k T \label{E_gamma}.
\end{eqnarray}

The interaction cross section of deuteron photo-disintegration can be well approximated by \citep{Bet50}:
\begin{eqnarray}
\sigma_{BL} = E1 + M1
\end{eqnarray}
where we express the electric dipole contribution
\begin{eqnarray}
E1= \frac{2}{3} ~\frac{e^2 ~\hbar~ \sqrt{B_d}~ (E_\gamma- B_d)^{\frac{3}{2}}}{c~ \epsilon_0 ~m_N~ E_\gamma^3~\left(1-\frac{r_t}{\hbar}\sqrt{m_N B_d}\right)}
\end{eqnarray}
and the magnetic dipole contribution
\begin{eqnarray}
M1 = \frac{e^2 \hbar(\mu_p - \mu_n)^2}{6~ \epsilon_0 ~m_N^2 ~c^3} \sqrt{\frac{B_d}{E_\gamma}-\left(\frac{B_d}{E_\gamma}\right)^2} \frac{\left(1-\sqrt{m_N B_d}\frac{a_s}{\hbar}+a_s(r_s+r_t)\frac{m_N B_d}{4 \hbar^2}-a_s(r_s - r_t)\frac{m_N (E_\gamma - B_d)}{4 \hbar^2} \right)^2}{(1+a_s^2\frac{m_N(E_\gamma-B_d)}{\hbar^2})(1-\frac{r_t}{\hbar} \sqrt{m_N B_d})}
\end{eqnarray}
in terms of $B_d$ and $E_\gamma$. $\epsilon_0$ denotes the electric constant, $a_s$ and $a_t$ singlet and triplet scattering length, $r_s$ and $r_t$ singlet and triplet effective range, $\mu_p$ and $\mu_n$ the magnetic moment of proton and neutron, respectively.\\
With this dependence of $\sigma_{BL}$ on the incident photon energy $E_\gamma$, we derive $\sigma_{(\gamma d \rightarrow np)}$ as the mean cross section of photo-disintegration per photon with $E_\gamma > B_d$:
\begin{eqnarray}
\sigma_{(\gamma d \rightarrow np)} = \frac{1}{n_{(\gamma>B_d)}} \frac{8 \pi}{(h c)^3}~\int_{B_d}^{\infty} ~E_\gamma^2~e^{-\frac{E_\gamma}{kT}}~\sigma_{BL}(E_\gamma)~ dE_\gamma.
\end{eqnarray}

We go back to eq. (\ref{gammas}) and divide by $\Gamma_{(\gamma d\rightarrow np)}$, in order to receive two terms, which we analyse separately:
\begin{eqnarray}\label{legammas}
\frac {\Gamma_{(np \rightarrow d\gamma)} }{\Gamma_{(\gamma d \rightarrow np)}} &=&
1 + \frac{\Gamma_{expansion}} {\Gamma_{(\gamma d \rightarrow np)}}.
\end{eqnarray}
We start with
\begin{eqnarray}
\frac {\Gamma_{(np \rightarrow d\gamma)} }{\Gamma_{(\gamma d \rightarrow np)}} &=&
\frac{n_n~\eta~\sqrt{\frac{8 k T}{\pi m_N}}~ \sigma_{(np \rightarrow d\gamma)}}{n_d~\left(1+\frac{n_n}{n_p} \right)~\frac{n_{\gamma}^*}{n_{\gamma}} ~c
~\sigma_{(\gamma d \rightarrow np)}}\\ \nonumber \\
&=& \frac{3.84 ~\eta~\sqrt{\frac{k T}{m_Nc^2}}~e^{\frac{B_d}{kT}}~ \frac{n_p}{n_d}~\frac{\sigma_{(\gamma e \rightarrow e \gamma)}}{\sigma_{(\gamma d \rightarrow np)}}}{\left(1+\frac{n_n}{n_p}\right)\left[\left(\frac{B_d}{kT}+1\right)^2+1\right]}~\frac{\sigma_{(np \rightarrow d\gamma)}}{\sigma_{(\gamma d \rightarrow np)}}
\end{eqnarray}
where $\sigma_{(np \rightarrow d\gamma)}$ is related to
$\sigma_{(\gamma d \rightarrow np)}$ by the detailed balance
\begin{eqnarray}
\frac{\sigma_{(np \rightarrow d\gamma)}}{\sigma_{(\gamma d \rightarrow np)}} \simeq
\frac{3 \langle E_\gamma \rangle^2}{2 m_N c^2(\langle E_\gamma \rangle - B_d)},
\end{eqnarray}
and $\langle E_\gamma \rangle$ is given by eq. (\ref{E_gamma}), leading to
\begin{eqnarray}\label{Term1}
\frac {\Gamma_{(np \rightarrow d\gamma)} }{\Gamma_{(\gamma d \rightarrow np)}}
&=& \frac{5.755 ~\eta}{1+\frac{n_n}{n_p}}\left(\frac{k T}{m_Nc^2}\right)^{\frac{3}{2}}e^{\frac{B_d}{kT}}~\frac{\sigma_{(\gamma e \rightarrow e \gamma)}}{\sigma_{(\gamma d \rightarrow np)}}~\left(\frac{n_n}{n_d}\right)^2\frac{n_p}{n_n}.
\end{eqnarray}
Next, we analyse the term on the right hand side of eq. (\ref{legammas}):
\begin{eqnarray}
\frac{\Gamma_{expansion}} {\Gamma_{(\gamma d \rightarrow np)}} &=& \frac{ \frac{n_d}{2 ~t}
}{
n_d~n_{\gamma}^* ~c
~\sigma_{(\gamma d \rightarrow np)}}\\ \nonumber \\  \label{Term2}
&=& 1.040~ \sqrt{\frac{G g_*(T_N)h^3}{c}}~ \frac{kT ~e^{\frac{B_d}{kT}}}{(B_d+kT)^2}~\frac{\sigma_{(\gamma e \rightarrow e \gamma)}}{\sigma_{(\gamma d \rightarrow np)}^2} ~\frac{n_p}{n_n}~\frac{n_n}{n_d}.
\end{eqnarray}

Both analysed terms depend on $T$ exponentially, but at $T=T_N$, when term(\ref{Term1}) reaches unity, term(\ref{Term2}) is still of order $10^{-2}$ and therefore negligible. In other words, the expansion rate of the universe is still dominated by the rate of deuteron photo-disintegration, when the primordial nucleosynthesis starts. Of course, the expansion of the universe causes the opening of the deuterium bottleneck, but its major influence is the reddening of the radiation, which enables the deuteron production to win over photo-disintegration. This simplifies our task drastically and we obtain the following equation for $T_N$, respectively $t_N$:
\begin{eqnarray}
 \frac{5.755 ~\eta}{1+\frac{n_n}{n_p}(T_N)}\left(\frac{k T_N}{m_Nc^2}\right)^{\frac{3}{2}}e^{\frac{B_d}{kT_N}}~\frac{\sigma_{(\gamma e \rightarrow e \gamma)}}{\sigma_{(\gamma d \rightarrow np)}}(T_N)~\left(\frac{n_n}{n_d}(T_N)\right)^2\frac{n_p}{n_n}(T_N) = 1
\end{eqnarray}
and finally the corresponding neutron to proton ratio:
\begin{eqnarray}\label{ratio_of_T_N}
\frac{n_p}{n_n}(T_N) = \frac{1}{X_n(T_f)} \exp{\left[\frac{1}{\tau_n} \sqrt{\frac{90
\hbar ^3 c^5}{32 \pi ^3 k^4 G}} \left(\frac{1}{\sqrt{g_*(T_N)}T_N^2}-
\frac{1}{\sqrt{g_*(T_f)}T_f^2}\right)\right]}-1
\end{eqnarray}
where $X_n$ is given by eq. (\ref{X_n}) .

The neutrinos have decoupled from equilibrium before the annihilation of electron positron pairs. Therefore the entropy due to this annihilation is transferred exclusively to the photons,
i.e. $g_*(T_N) = 2 +~ \frac{7}{8}~ 2~ N_\nu~ (\frac{4}{11})^{\frac{4}{3}}$.

Assuming neutron conservation after $t_N$, we finally calculate $Y_P$, the primordial $^4$He abundance by weight. Since $^4$He is not further transformed into heavier
nuclei, because elements with nucleon mass number A=5 and A=8 are insufficiently stable to
function successfully as intermediate products for nucleosynthesis at the available
densities, we derive:
\begin{eqnarray}\label{Y_p}
Y_P &=&  \frac{\frac{1}{2} n_n m_{He}}{\frac{1}{2} n_n m_{He} + (n_p - n_n) m_p} = \frac{1}{1+2 \frac{m_p}{m_{He}} \left(\frac{n_p}{n_n}(T_N)-1 \right)}
\end{eqnarray}
where $m_{He}$ denotes the mass of the helium nucleus and $\frac{n_p}{n_n}(T_N)$ is given by eq. (\ref{ratio_of_T_N}).

This analytic expression for $Y_P$ reproduces the observation based results and the numerical results (see section results for details), assuming neutron conservation when $52 - 66$ \% of the neutrons have reached stable states inside light nuclei. Therefore eq. (\ref{Y_p}) provides our basis for finding the
dependence of $Y_P$ and the possible deviation of $v$ from its present value $v_0$, in
order to finally constrain $v\over v_0$ by recent results on the primordial $^4$He abundance.

Expressing all key-parameters of $Y_P$ by terms of $v$, we start with the most important one, the deuteron binding energy. Within our narrow range of interest ($|\frac{v-v_0}{v_0}|< 0.5$\%), we determine its dependence on $v$ by varying the pion mass in different nucleon-nucleon-potential-model calculations, based on \citep{Are91}, and derive:
\begin{eqnarray}
B_d(v) \simeq B_d(v_0)~(A- (A-1) \sqrt{v/v_0}),
\end{eqnarray}
where $A$ is a model dependent constant as follows:\\
\begin{tabular}{llr}
Bonn-A potential \citep{MaH87}:  &    $A~=$&$2.3$,\\
Paris potential \citep{LaL80}:   &    $A~=$&$28.2$,\\
Argonne V14 potential \citep{WiS84}:& $A~=$&$61.4$.\\
\end{tabular}\\
\\
For our purpose, we take the mean average $A= 30.6$.

As $B_d$ changes, $E_\gamma$ and the cross sections $\sigma_{(\gamma d
\rightarrow np)}$ and $\sigma_{(np \rightarrow d\gamma)}$ change, accordingly. Furthermore, we have to consider, that the mass of the electron varies proportionally
\begin{eqnarray}
m_e(v) = m_e(v_0)~\frac{v}{v_0},
\end{eqnarray}
which enters the Klein-Nishina cross section.

Concerning $\tau_n$, the mean lifetime of the free neutron, we use the expression \citep{Gas08}, based on \citep{Mue04}:
\begin{eqnarray}
\tau_n(v) \simeq \tau_n(v_0)~( 1 - 4.88 ~\frac{v-v_0}{v_0}).
\end{eqnarray}

Next, we have to consider the change on $Q$, the neutron to proton mass difference, which influences the freeze-out concentration. We separate the electromagnetic contribution \citep{Gas82} and obtain
\begin{eqnarray}
Q \simeq (-0.76 + 2.0533317 ~\frac{v}{v_0})~ [MeV].
\end{eqnarray}

The Fermi coupling constant $G_F$ is related to $v$ by \citep{Dixit}:
\begin{eqnarray}\label{G_F}
G_F(v) = \frac{1}{v^2 \sqrt{2}}~ \left[\frac{1}{GeV^2}\right].
\end{eqnarray}

Finally, we derive a relation
between $Y_P$ and $v$, to constrain the permitted variation of the Higgs vacuum
expectation value by the primordial $^4$He abundance:
\begin{eqnarray}\label{result1}
\Delta Y_P &\simeq& - 38~ \left(\frac{\Delta v}{v}\right)^2 - 2.08~ \left(\frac{\Delta v}{v} \right) + \nonumber \\ \nonumber \\ &+& 0.0355 ~ \left(\frac{N_\nu-3}{3} \right) + 0.0874~ \left(\frac{\Delta G}{G} \right) + 0.0042~\ln \left(\frac{\Delta \eta}{\eta}+1 \right).
\end{eqnarray}
Varying each parameter separately (assuming the others fixed), we derive:
\begin{eqnarray}\label{result2}
\Delta Y_P &\simeq& 0.106~ \left(\frac{\Delta B_d}{B_d}\right) + 0.056 ~ \left(\frac{\Delta \tau_n}{\tau_{n}} \right) - 0.235~\left(\frac{\Delta G_F}{G_{F}} \right)
- 0.352~\left(\frac{\Delta Q}{Q} \right) - \nonumber \\ \nonumber \\ &-& 0.006~\left(\frac{\Delta m_e}{m_e} \right) - 0.195~\left(\frac{\Delta g_A}{g_A} \right) + 0.0355~ \left(\frac{\Delta N_\nu}{3} \right) + \nonumber \\ \nonumber \\ &+& 0.0874~ \left(\frac{\Delta G}{G} \right) - 0.2602~ \left(\frac{\Delta \hbar}{\hbar} \right)+ 0.0042~\ln \left(\frac{\Delta \eta}{\eta}+1 \right).
\end{eqnarray}
We briefly note, that using eq. (\ref{result1}), one can also obtain constraints on $N_\nu$, $G$, $\hbar$ and $\eta$.

\section{Results}
Constraining the possible time variation of $v$, we use observation based results as well as the standard big bang nucleosynthesis code, developed by \cite{Wag73} and \cite{Kaw92}. This standard code still seems adequate for our purpose,  although newer nuclear reaction rates have been evaluated \citep{Des04}.\\
The baryon to photon ratio is given by $\eta = (273.9 \pm 0.3)~10^{-10} ~\Omega_b h^2$ \citep{Ste06}, where $\Omega_b$ is the present ratio of the baryon mass density to the critical density and $h$ is the present value of the Hubble parameter in units of $100$ kms$^{-1}$Mpc$^{-1}$. We take $100~ \Omega_b h^2 = 2.273 \pm 0.062$, the 5-year mean value of WMAP \citep{Kom08}, and obtain:
\begin{eqnarray*}
6.049 \cdot 10^{-10} \leq ~\eta ~ \leq ~6.403 \cdot 10^{-10}.
\end{eqnarray*}
Implementing $\eta$ and $\tau_n = 878.5$ s \citep{Serebrov}, the numerical code delivers
\begin{eqnarray*}
Y_1 = 0.2467 \pm 0.0003
\end{eqnarray*}
and thus eq. (\ref{result1}) constrains the possible time variation of $v$:
\begin{eqnarray*}
\left|\frac{\Delta v}{v}\right| ~ \leq ~1.5 \cdot 10^{-4}.
\end{eqnarray*}
Using the observation based results of \cite{Izo07} we derive
\begin{eqnarray*}
Y_2 = 0.2516 \pm 0.0011~~\Rightarrow~~
\left|\frac{\Delta v}{v}\right| ~ \leq~ 5.6 \cdot 10^{-4}.
\end{eqnarray*}
and the results of \cite{Pei07} lead to
\begin{eqnarray*}
Y_3 = 0.2477 \pm 0.0029 ~~\Rightarrow~~
\left|\frac{\Delta v}{v}\right| ~ \leq~ 1.4 \cdot 10^{-3}.
\end{eqnarray*}
Combining the observations of H II regions ($Y_2$ and $Y_3$) and the numerical simulation based on WMAP results ($Y_1$), according to
\begin{eqnarray}
\Delta{Y}&=& Y_P(\textsf{H II regions}) - Y_P(\textsf{WMAP + SBBN}),
\end{eqnarray}
we alternatively derive (using eq. (\ref{result1})) two more conservative estimates: 
\begin{eqnarray}
Y_2 - Y_1 &=& 0.0049 \pm 0.0011 ~~\Rightarrow~~ \left|\frac{\Delta v}{v}\right| ~ =~ (2.4 \pm 0.6) \cdot 10^{-3} \nonumber \\ \nonumber \\
Y_3 - Y_1 &=& 0.0010 \pm 0.0029 ~~\Rightarrow~~ \left|\frac{\Delta v}{v}\right| ~ =~ (0.5 \pm 1.4) \cdot 10^{-3}. \nonumber
\end{eqnarray}

We avoid the term "observational results" because the cited publications more or less
consist of interpretations of the observational $^4$He abundance plus theoretical input and
constraints by the cosmic microwave background. The different interpretation
as a result of the badly understood systematics lead to incompatible data. For consistency, we only cite data based on recent He I recombination coefficients by \cite{Por05,Por07}.

\section{Conclusions}
Big-bang nucleosynthesis offers the deepest reliable probe of the early universe. Its predictions of the light element abundances play a major role in constraining cosmological models. The increasing precision of observational results on primordial abundances opens new scientific fields that can be tested by BBN.
We present a calculation, how observations on primordial $^4$He may provide insight into the fundamental property of elementary particles: the Higgs vacuum expectation value. We find constraints on its allowed time variation $\left|\frac{\Delta v}{v}\right| ~ \leq 1.5 \cdot 10^{-4}$.

\acknowledgments

We thank D. Tekle and A. Bauer for help on the Kawano code and S. Winitzki and A. Jessner for valuable comments on the manuscript.
\appendix

\clearpage

\end{document}